\documentclass[aps,pre,showpacs,floatfix,superscriptaddress,twocolumn]{revtex4}
\usepackage{graphicx}
\usepackage{amsmath}
\usepackage{amsfonts}
\usepackage{epstopdf}

\begin{document}


\title{A kinetic model and scaling properties for non-equilibrium clustering of self-propelled particles}

\author{Fernando Peruani} 
\affiliation{Laboratoire J.A. Dieudonn{\'e}, Universit{\'e} de Nice Sophia Antipolis, UMR 7351 CNRS, Parc Valrose, F-06108 Nice Cedex 02, France}
\author{Markus B\"ar} 
\affiliation{Physikalisch-Technische Bundesanstalt, Abbestrasse 2-12, 10587 Berlin, Germany}

\date{\today}

\begin{abstract}
%
We demonstrate that the clustering statistics and the corresponding phase transition to non-equilibrium clustering  found in many experiments and simulation studies with  self-propelled particles (SPPs) with alignment
 can be obtained from a simple kinetic  model. The key elements of this approach are the scaling  of the cluster cross-section with the cluster mass -- characterized by an exponent $\alpha$ -- 
and the scaling of the cluster perimeter with the cluster mass -- described by an exponent $\beta$. 
The analysis of the kinetic approach reveals that the SPPs exhibit two phases: 
i) an individual phase, where the cluster size distribution (CSD) is dominated by an exponential tail that defines a characteristic cluster size, 
and ii) 
a collective phase characterized by the presence of non-monotonic CSD 
with a local maximum at large cluster sizes. 
At the transition between these two phases the CSD is well described by a power-law with a critical exponent $\gamma$, which is a function of $\alpha$ and $\beta$ only. The critical exponent  is found to be in the range 
$0.8 < \gamma < 1.5$ in line with observations in experiments and simulations. 
%
%
\end{abstract}
\pacs{87.18.Gh, 05.65.+b, 87.18.Hf}


\maketitle

\section{Introduction}

Many experimental self-propelled particle (SPP) systems~\cite{vicsek2012, marchetti2012c}, from actin filaments driven by molecular motors~\cite{schaller2010, kohler2011}, gliding and swimming bacteria~\cite{zhang2010, peruani2012, starruss2012}, to 
active colloidal particles~\cite{theurkauff2012, palacci2013} exhibit a remarkably rich cluster dynamics.
%
%
In particular,  the formation of large moving polar clusters has been observed in most of these examples. Such clusters are formed by particles that move roughly in  the same direction. 
 %
Simulations of simple models of SPPs have revealed that similar clustering dynamics are observed for  SPPs with both, polar~\cite{vicsek1995, huepe2004, romensky2013} and nematic~\cite{peruani2008, ginelli2010, peruani2010} 
alignment interactions. 
More detailed models display similar clustering statistics, e.g., in    
simulations of self-propelled rods~\cite{peruani2006, yang2010, weber2012}, self-propelled disks~\cite{grossman2008, deseigne2008}, 
 particles with hydrodynamic coupling~\cite{llopis2006}, 
and swimming particles with flagella, e.g., in sperm cells~\cite{yang2008}. 
%

\begin{figure}
\centering
\resizebox{0.95\columnwidth}{!}{\rotatebox{0}{\includegraphics{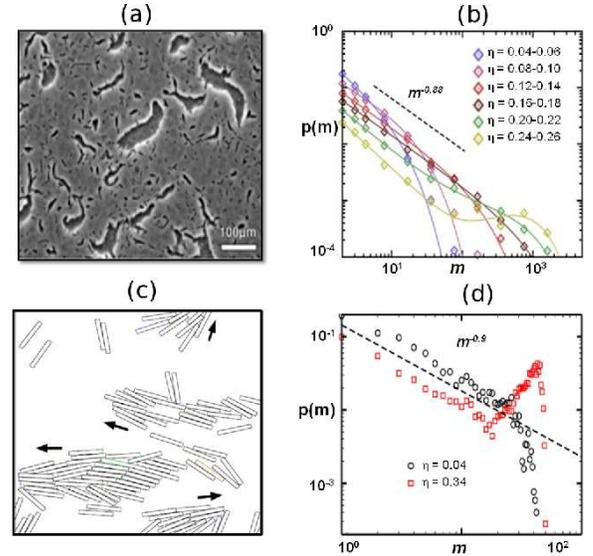}}}
\caption{Examples of non-equilibrium cluster formation in  self-propelled particle systems. (a) Large moving clusters in experiments with myxobacteria, where it has been found that the cluster size distribution  (CSD) is a function of the cell density as shown in (b) (see~\cite{peruani2012}).  
(c) Snapshot of moving clusters in self-propelled rod simulations, where the CSD is known to be function of the particle density, as illustrated in (d) (see~\cite{peruani2006}). 
} \label{fig:illustration}
\end{figure}
%

\begin{figure}
\centering
\resizebox{0.99\columnwidth}{!}{\rotatebox{0}{\includegraphics{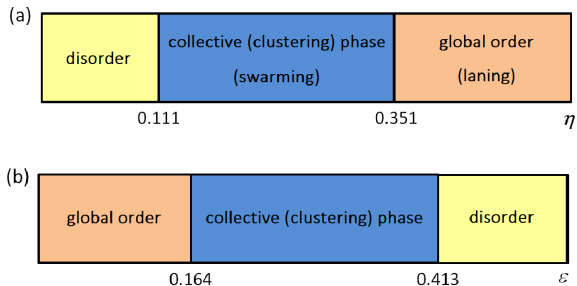}}}
\caption{Examples for the onset of the collective (clustering) phase (blue) and global order (orange) in simulations. (a) Self-propelled hard rods in two-dimensions with aspect ratio $\kappa = 13$, $\nu$ denotes the fraction of covered area~\cite{wensink2012, wensink2012b}. (b) Modified Vicsek model with nematic alignment with density $\rho = 0.25$~\cite{peruani2008}. The noise value for the transition to global order was taken from simulation with $> 10^6$ self-propelled particles~\cite{ginelli2010}, while the noise value for the onset of clustering in a system with $16.384$ self-propelled particles is based on the results of~\cite{peruani2010}.
} \label{fig:struct}
\end{figure}

In many of these systems, a peculiar phase, characterized by the existence of remarkably large moving clusters, has been observed. 
One refers to this phase of collective motion as non-equilibrium clustering. 
This phase  appears often as an intermediate phase between a completely disordered phase, with homogeneous density, and a phase with global orientational order. 
An experimental case (myxobacteria)  and a simulation example (self-propelled rods) are shown in 
Fig.~\ref{fig:illustration}. The occurrence of large clusters is strongly correlated with a crossover in the shape of the cluster size distributions (see Fig. ~\ref{fig:illustration}b, d). 
Figure \ref{fig:struct}  illustrates recent findings in simulations of self-propelled hard rods~\cite{wensink2012, wensink2012b} and in simulations with a modified Vicsek model with nematic alignment~\cite{peruani2008, ginelli2010, peruani2010}. The collective clustering phase is 
characterized by a non-monotonic cluster size distribution (CSD) with a characteristic peak at large cluster sizes, see CSD for high densities in Fig.~\ref{fig:illustration}.  
At the onset of the collective clustering phase, the CSD follows a power law with a characteristic exponent. 
It has been observed in SPP experiments~\cite{zhang2010, peruani2012} and simulations~\cite{peruani2006, yang2010} 
that the CSD can be power-law distributed with the exponent typically varying  between $0.85$ and $1.35$.
While the range of exponents  indicates the absence of a  universal scaling at the onset of the collective clustering phase, it is not yet clear what determines the actual value of the exponent.  

The examples displayed in Figs.~\ref{fig:illustration} and \ref{fig:struct} indicate that the collective clustering phase often occurs in the absence of global order. Other models, e.g. the classical Vicsek model display a similar phase, but in the presence of global order~\cite{huepe2004, chate2008, romensky2013}. A word of caution is in place here. Most investigations that focused on the cluster statistics (both, in experiments and simulations) were carried out with small to intermediate numbers (around $10^2 - 10^3$) of self-propelled particles, while investigations regarding the onset of global order were conducted with rather large numbers ($> 10^5$). Hence, it is not completely clear if and how the system size, respectively the particle number, affects the onset of this collective clustering phase.

%

%
Here, we focus on the emergence of the collective clustering phase in the absence of global order, i.e. we assume that cluster velocities are uncorrelated.  
%
%
%
We propose a kinetic clustering theory, based on the approach presented by us in~\cite{peruani2006}, and extend our earlier treatment to coagulation and fragmentation kernels that depend, respectively,  
on the scaling of the cluster cross-section with cluster "mass", i.e. number of particles in the cluster,  and the scaling of cluster perimeter with cluster mass. 
The scaling of the cluster cross-section and cluster perimeter are characterized by exponents $\alpha$ and $\beta$, respectively.  
Furthermore, we present  a comprehensive  analysis of this kinetic cluster model, including a finite size (FS) study of it. The FS  study of the kinetic model reveals that:
i) the transition to the collective clustering phase -- characterized by a non monotonic CSD -- is generic to all SPPs with either a polar or nematic alignment mechanism, 
and ii) that there exist a critical asymptotic CSD exponent $\gamma$ which is function of $\alpha$ and $\beta$ only. By studying the physically meaningful parameter 
space $\alpha$-$\beta$ we find that $\gamma$ always falls in the range $0.8 < \gamma < 1.5$. 
Notice that in the absence of cluster-cluster correlations,  this transition marks the onset  of collective motion as observed in~\cite{peruani2012}. 
Though  the simplified clustering theory is strictly speaking only valid in the absence of global order, we show through simulations that a comparable picture holds for systems where global order is observed.

%
%

The paper is organized as follows. In section~\ref{sec:clusteringTheory} we present our simple kinetic clustering model under the assumption that there are no cluster-cluster correlations
and show that for any finite system there are two clustering phases. 
We perform a system size analysis of the kinetic model in Sec.~\ref{sec:scalingProperties}, where we report on the scaling properties 
of the obtained cluster size distribution (CSD) which we find to depend on the scaling of the cluster cross-section and cluster perimeter.  
In Sec.~\ref{sec:scalingProperties} we study cluster formation in a SPP model that is known to exhibit cluster-cluster correlations and contrast the obtained results 
with our cluster-cluster uncorrelated clustering theory 
The implications and limitations of the proposed kinetic approach are discussed in Sec.~\ref{sec:concludingRemarks}.


\section{A kinetic model for clustering}\label{sec:clusteringTheory}  
%
%
We look for a description of the clustering process in terms  of $\langle n_1(t)\rangle, \langle n_2(t) \rangle, ..., \langle n_N(t) \rangle$, 
where $\langle n_i(t)\rangle$ represents the average value at time $t$ of clusters formed by $i$ particles, and $N$ the number of particles in the system.  
To ease the notation, we refer to $\langle n_i(t) \rangle$ as $n_i(t)$. 
Our strategy consists on deriving a coagulation Smoluchowski equation with fragmentation for these $n_i(t)$ objects. 
For an introduction to this kind of coagulation equations we refer the reader to~\cite{redner}.  
In SPP systems, the only conserved quantity is the number of particles $N$. 
Neither the overall orientational order nor the number of clusters are conserved. Thus, our generalized coagulation Smoluchowski equation should conserve the number of particles. 
We start by simplifying the clustering dynamics. 
Since we are dealing with self-propelled particles, 
we assume that clusters are the result of either an explicit or an effective 
alignment mechanism such that particles inside a cluster move coherently in the same direction. 
This implies that clusters move at  
speeds comparable to that of the individual particles, independently of the size of the cluster. Below we also discuss what can be expected if this condition is relaxed.  
On the contrary, clusters of passive particles,  driven by thermal fluctuations, are such that their 
mobility decreases as function of their size. 
We stress that the assumption of a size-independent cluster speed evidences  the non-equilibrium nature of our simple cluster theory. 
%
%
We further assume that (binary) collisions among clusters may result in cluster-cluster 
fusion, and neglect the possibility of cluster fragmentation induced by
cluster-cluster collision. 
This assumption is justified as long as the dynamics is overdamped, and is in
accordance to what is observed in  experiments with
gliding bacteria~\cite{peruani2012}, and simulations with self-propelled rods~\cite{peruani2006}.
We simplify the cluster fragmentation dynamics by assuming an evaporation-like
process by which clusters shrink in size by loosing  one by one those particles that are on the cluster boundary. 
More complex fragmentation process can be ignored as exponentially unlikely
events. 
%
%
%
%
In summary, we are assuming an irreversible clustering process in which a
cluster of mass $j$ can undergo the following ``reactions'':
\begin{eqnarray}\label{eq:reactions}
\begin{array}{cc}
C_{j} + C_{k} \stackrel{A_{j,k}}{\rightarrow}  C_{j+k} , & 
C_j  \stackrel{B_{j}}{\rightarrow}  C_{j-1} + C_{1} \,. 
\end{array}
\end{eqnarray}
Notice that since the reaction $C_{j+k} \to C_j + C_k$ does not occur, the
process is, in this sense, irreversible.
Now, we look for a description of the process in terms of the (average)  number $n_j(t)$ of clusters
with $j$ particles (i.e., number of $C_{j}$'s) at time $t$, whose time evolution takes the form:
\begin{eqnarray}
\dot{n}_{1}&=&2B_{2}n_{2}+\sum_{k=3}^{N}B_{k}n_{k}-\sum_{k=1}^{N-1}A_{k,1}n_{k}n_{1}
\nonumber \\
\dot{n}_{j}&=&B_{j+1}n_{j+1}-B_{j}n_{j}-\sum_{k=1}^{N-j}A_{k,j}n_{k}n_{j}
\nonumber \\
&&+\frac{1}{2}\sum_{k=1}^{j-1}A_{k,j-k}n_{k}n_{j-k} \ \quad \mbox{for} \quad j = 2, .....,N-1
\nonumber \\
\dot{n}_{N}&=&-B_{N}n_{N}+\frac{1}{2}\sum_{k=1}^{N-1}A_{k,N-k}n_{k}n_{N-k} \label{rea_vicsek}
\end{eqnarray}
where the dot denotes the time derivative, $B_{j}$ represents the
 rate at which a cluster of mass $j$ looses particles,  defined as 
\begin{eqnarray}\label{eq:rate_splitting}
B_{j}= \frac{D}{d^2}\,j^{\beta} \, ,
\end{eqnarray}
and $A_{j,k}$ is the collision rate
between clusters of mass $j$ and $k$, defined by
\begin{eqnarray}\label{eq:rate_collision}
A_{j,k}= \frac{q_s v_c \sigma_0}{L^2}\left(j^{\alpha}+ k^{\alpha}\right) \, ,
\end{eqnarray}
where $L^2$ is the area of the two-dimensional space where particles move and $v_c$ is the cluster speed, which we assume to be 
$v_c \sim v_0$ with $v_0$ the speed of individual particles.  
Notice that Eqs.~(\ref{rea_vicsek}) are such that $\sum m\,\dot{n}_m=0$, and
thus the number of particles $N=\sum m\,n_m(t)$ is conserved. 
In Eq.~(\ref{eq:rate_splitting}), $d^2/D$ is the typical time a particle located on the 
 cluster boundary needs to detach from a cluster, with 
$d$ the  maximum distance two particles can be apart to be still considered as
{\it connected} and $D$ the diffusion coefficient with respect the center of mass of 
 the cluster. 
%
%
 We stress that in Eq.~(\ref{rea_vicsek}) we have assumed that cluster are uncorrelated. 
Several statistical features of the model can be given explicitly, if one assumes that SPP follow a dynamics where their translation is determined by a constant speed and the direction of motion is subject to alignment interactions with neighboring SPP and an angular noise with amplitude $\eta$. An example of such dynamics is given by the following equations of motion: 
\begin{eqnarray}\label{eq:mot_theta}
\theta^{t+1}_j&=& \arg\left[ \sum_k f\left(e^{i\theta^t_k}, e^{i\theta^t_j} \right)\right] + \eta^t_j \,, \\
\nonumber
 \mathbf{x}^{t+1}_j &=& \mathbf{x}^{t}_j + v_0 e^{i \theta^{t}_k} \, ,  
\end{eqnarray}
where $j$ is the index of the particle, $\theta^{t}_j$ defines (in two dimension) the particle moving direction, $\mathbf{x}^t_j$ denotes the position at time $t$, the sum is taken over all particles within a unit distance of $j$, and $\eta^t_j$ is a uniformly distributed random variable such that  $-\eta/2<\eta^t_j<\eta/2$, 
 $\langle \eta^t_j \rangle=0$, and $\langle \eta^t_j \eta^{t'}_k \rangle=(\eta^2/12) \delta_{t,t'} \delta_{j,k}$. 
The dynamics given by Eq.~(\ref{eq:mot_theta}) is 
 frequently used in simulation studies, e.g. in~\cite{vicsek1995, huepe2004, peruani2008,  baglietto2009, ginelli2010, peruani2010, peruani2011b, romensky2013} (for a review, see~\cite{vicsek2012}).  
More specifically, $f\left(e^{i\theta^t_k}, e^{i\theta^t_j} \right)=e^{i\theta^t_k}$ defines a polar alignment rule as in the Vicsek model~\cite{vicsek1995}, 
$f\left(e^{i\theta^t_k}, e^{i\theta^t_j} \right)=\text{sign}[ \cos(\theta^t_k- \theta^t_j)] e^{i\theta^t_k}$ defines SPP with a nematic alignment rule as used~\cite{peruani2008, ginelli2010}.
Assuming that the SPP obey Eq.~(\ref{eq:mot_theta}), the diffusion around the center of mass of a polar cluster of SPP  
can be approximated, as detailed in~\cite{peruani2010}, by 
 $D = v_0^2 [1- (\sin(\eta/2)/(\eta/2))^2] \Delta t$, where  $v_0$ is the speed of individual particles, and $\Delta t$ sets the discrete time step for a typical SPP dynamics such as the one given by Eq.~(\ref{eq:mot_theta}).  
By making a Taylor expansion we find that $D \sim v_0^2/12 \eta^2 \Delta t$.  
Notice that for non-interacting random walkers subject for the same kind of noise, $D \propto v_0^2 / \eta^2$. 
In addition, we have to consider that the splitting rate $B_{j}$ has to be proportional to the number of
particles on the cluster boundary.  
In Eq.~(\ref{eq:rate_splitting}), we have  assumed that the perimeter $l$ of a cluster scales with its mass $j$  as
$l \propto j^{\beta}$, with $\beta$ constrained to $1/2\leq\beta\leq1$. 
Let us provide a physical context to this assumption, always assuming we are in a two dimensional space.  
Surface tension would tend to minimize cluster perimeters, and so clusters would be round and $\beta = 1/2$,  as observed in liquid-vapor drops~\cite{stauffer}. 
On the other hand, for random deposition of particle as in classical percolation,  $\beta \approx 1$. 
If clusters are  ``chains'' of particles then also  $\beta=1$.  
In short, we expect $0.5 < \beta < 1$.   
%
%
%
%
\begin{figure}
\centering
\resizebox{\columnwidth}{!}{\rotatebox{0}{\includegraphics{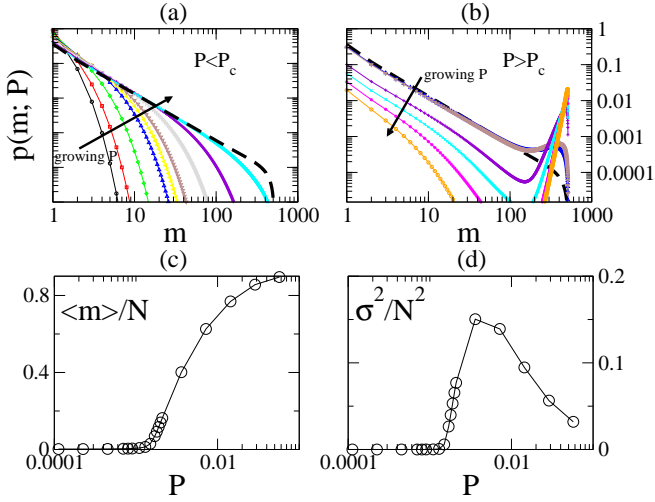}}}
\caption{For a given system size $N$, the cluster size distribution (CSD) $p(m)$ is
  monotonically decreasing for $P<P_c$ (a), while for $P>P_c$ the distribution
  exhibits a peack at large cluster sizes (b). At $P=P_c$ the distribution
  $p(m)$ is a power-law.
The transition between these two behaviors, (a) and (b), can be represented in
a more phase transition-like form, by looking at first two moments of the
distribution $p(m)$, (c) and (d), respectively. Other parameters: $N=512$ and $\beta=1/2$.} \label{fig:scaling_at_Pc}
\end{figure}
%
%
%
\begin{figure}
\centering
\resizebox{0.7\columnwidth}{!}{\rotatebox{0}{\includegraphics{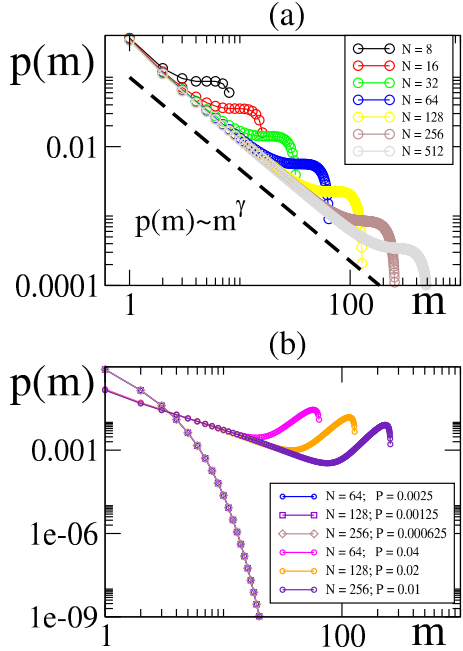}}}
\caption{(a) Scaling of $p(m)$ at the critical point $P_c(N)$ for various
  system sizes $N$. Notice that for $N\to\infty$ the distribution is given by Eq.~(\ref{eq:pm_at_Pc}). 
(b) Below and above $P_c(N)$, $p(m)$ exhibits the scaling given by
Eq.~(\ref{eq:pm_at_Pc}). (a) and (b) corresponds to $\alpha = \beta=1/2$. For the value
of $\gamma$ see Fig.~\ref{fig:exponents}.} \label{fig:exponents_0p5}
\end{figure}
%
%

The collision rate $A_{j,k}$ is derived in analogy to the collision rate in kinetic gas theory~\cite{reif} , 
which means we assume clusters move ballistically in between collisions, which leads to a collision rate 
 proportional to the sum of the scattering cross section of the involved clusters, the relative cluster speed, and cluster density. 
Due to the latter assumption,   $A_{j,k} \propto L^{-2}$.  
The relative cluster speed can be assumed to be not too different from that of the individual clusters, $v_c$. 
In SPP systems, we find often that individual cluster speed is close to that of individual particles, i.e., $v_c \sim v_0$. 
If $v_c$ exhibits a dependency on the cluster size of the form $v_c \propto j^{-\xi}$, the exponent $\xi$ can be absorbed into $\alpha$. 
For simplicity, here we assume that $v_c \sim v_0$. 
For instance, the speed of SPP clusters can be approximated by $v_c = v_0 \sin(\eta/2)/(\eta/2) \propto v_0 (1- \eta^2)$~\cite{peruani2010}.
%
%
The scattering cross-section of a 2D cluster is the (average) projection of the cluster on a given axis. 
Thus, scattering cross section has to be always less or equal to the cluster perimeter. 
If the scattering cross section of a cluster scales as $\propto
\sigma_0\, j^{\alpha}$, the previous observation implies that $\alpha \leq \beta$ is the only meaningful physical scenario.  
Below, we will see that there are two qualitatively very different physical scenarios, $\alpha = \beta$ and  $\alpha < \beta$. 
Finally,  $q_s$ represents the probability that a cluster-cluster collision resulted in a successful fusion of clusters. 
Here, we assume it is a constant but certainly it can depend on various (intensive) variables of the actual system. 
In particular, it depends on the alignment symmetry. A very rough assumption would be to assume that if $q_s=q_0$  for the ferromagnetic alignment, $q_s=q_0/2$  for the nematic one. 
Eqs.~(\ref{rea_vicsek})  are scaled and transformed into a dimensionless form  by dividing Eq.~(\ref{eq:rate_splitting})
and~(\ref{eq:rate_collision}) by $D/d^2$, which leads to the following dimensionless parameter: 
\begin{eqnarray}\label{eq:P_def}
P=\frac{q_s \sigma_0 d^2 v_c}{L^2 D}\,.
\end{eqnarray}
Thus, Eq.~(\ref{eq:rate_splitting}) reduces to $B_j = j^{\beta}$, and
Eq.~(\ref{eq:rate_collision}) to $A_{j,k}=P(j^{\alpha}+k^{\alpha})$. 
The parameter $P$ controls the relative weight of fragmentation with
respect to coagulation. 
For model type considered here, we obtain $P \propto (L \eta)^{-2}$. 
The computation of this critical point $P$ can be done by directly 
studying the stability of the individual phase, and ignoring the actual shape of the CSD, as recently proposed in~\cite{weber2013}.

%
\begin{figure}
\centering
\resizebox{\columnwidth}{!}{\rotatebox{0}{\includegraphics{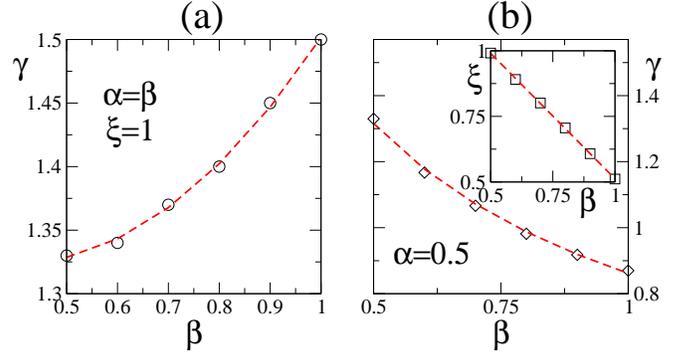}}}
\caption{
Exponents $\gamma$ as function of the exponents $\alpha$ and $\beta$. 
In (a) $\alpha=\beta$, while in (b) $\alpha=0.5$ and $\beta>\alpha$. 
The inset in (b) shows that when $\beta > \alpha$, $\xi<1$; for $\alpha=\beta$, $\xi =1$, see Eq. (\ref{eq:P_c}).  } \label{fig:exponents}
\end{figure}

From direct numerical integration of Eqs.(\ref{rea_vicsek}), in its
dimensionless version, with initial
condition $n_j(t=0)=N\,\delta_{j,1}$, we find that the (weighted) 
cluster size distribution (CSD), defined as 
\begin{eqnarray}\label{eq:csd_def}
p(m,t) = \frac{m\,n_m(t)}{N}\, ,
\end{eqnarray}
reaches a steady state, i.e., $p(m,t\to\infty)=p(m)$. 
This also implies that the number of cluster in the systems, $M(t)=N\sum
m^{-1}p_m(t)$, also reaches a steady state value as $t\to\infty$. 
In the literature the term CSD is frequently used to refer to
$\tilde{p}(m,t)=n_m(t)/M(t)$. 
The disadvantage of this definition is that its normalization constant,
$M(t)$, varies with time. 
While $\tilde{p}(m)$ refers to the probability of finding a cluster of size
$m$, $p(m)$ indicates the probability of a randomly selected particle to be in a cluster of size $m$.
Given $\alpha$ and $\beta$, the CSD $p(m)$ depends on $N$ and the value of $P$. 
Fig.~\ref{fig:scaling_at_Pc} summarizes the clustering behavior with $P$ for a given $N$. 
Notice there exists a critical $P_c(N)$ that separates two different clustering behaviors. 
For small values of $P$, the distribution $p(m)$  is monotonically decreasing, dominated by an exponential tail. 
As  $P \to P_c$, $p(m)$ approaches a power-law, with a system size cut-off.  
For $P>P_c$, $p(m)$ is non-monotonic and exhibits a peak at large cluster sizes emerges. 
This dramatic change of behavior at $P_c$ unveils a phase transition, which 
is evidenced by Fig.~\ref{fig:scaling_at_Pc}(c) and (d) that show 
$\langle m \rangle = \sum m \,p(m)$  and $\sigma^2 = \sum (m - \langle m \rangle)^2 \, p(m)$, respectively. 
%
%
Below, we will see that this transition is a genuine phase transition -- in the thermodynamical sense -- only for $\alpha=\beta$.

\section{Scaling properties} \label{sec:scalingProperties}
For every $N$, we numerically estimate $P_c(N)$ as the point at which $p(m)$ is no longer
monotonically decreasing. 
At the critical point $p_c(N)$, $p(m)$ scales as:
\begin{eqnarray}\label{eq:pm_at_Pc}
p(m; P_c) \propto m^{-\gamma}  \, ,
\end{eqnarray}
where $\gamma$ is a critical exponent that depends on $\alpha$ and $\beta$. 
This fact is illustrated by Fig.~\ref{fig:scaling_at_Pc}(a) that  shows that $p(m; P_c)$  follows the scaling given by Eq.~(\ref{eq:pm_at_Pc}) up to a given
cluster size $m^*(N)$ above which the finite size of the system becomes evident. 
We find the in general, below the critical point, the following scaling is obeyed by $p(m)$: 
\begin{eqnarray}\label{eq:scaling_p}
p(m;N, P, \alpha, \beta) = K\, p(m; bN, b^{-\xi}P,\alpha, \beta) \, ,  
\end{eqnarray}
where $K$ is a constant, see collapse of the curve in Fig.~\ref{fig:exponents_0p5}(b).    
For $\alpha = \beta$, $K=1$, while for $\alpha \neq \beta$, $K$ 
depends on $b$, $N$, and $\beta$. 
This means that if we double the 
system size and reduce $P$ by $2^{-\xi}$, we fall on the same 
distribution $p(m)$ for $\alpha = \beta$, and for $\alpha<\beta$ there is a multiplicative constant $K$.  
For $P< P_c$, we find that the CSD is well fitted by $p(m) \sim m^{-\gamma} \exp(-m/\tilde{m}(P)) $.

For  $P> P_c$, we observe that there is a shift with $N$ of the peak that emerges at large cluster sizes, Fig. \ref{fig:scaling_at_Pc}(b)
The scaling given by Eq.~(\ref{eq:scaling_p}) works up to a given size $m_m(N)$ which is given by the minimum of $p(m)$. 
We are interested in knowing the behavior of the peak with the system size $N$. 
%
In order to answer these questions we study the scaling of $Q(N)=\sum_{m_{m(N)}}^{\infty} p(m)$. 
We find that for $\alpha=\beta$, $Q(N\to\infty) \to 1$. 
This indicates the presence of a collective phase in which most of the particles are part of large clusters. 
Notice that this does not mean the formation of a single giant cluster; the width of the peak does not shrink to zero. 
For $\alpha<\beta$,  $Q$ does not converge to $1$ as $N \to \infty$. 
This is because for $\alpha<\beta$, in the limit of $N\to\infty$ the transition does not occur.  
Nevertheless, for any finite $N$, we always observe the  transition to the collective phase.  
This observation can be understood by studying the system size behavior of $p_c(N)$.  
We find that: 
\begin{eqnarray}\label{eq:P_c}
P_c(N) \propto N^{-\xi} \, ,
\end{eqnarray}
with $\xi=1$ for $\alpha=\beta$, and $\xi<1$ for $\alpha<\beta$. 
For a fixed value of $\alpha$, $\xi$ becomes a function of $\beta$ as illustrated in the inset of 
Fig. \ref{fig:exponents} (b) for $\alpha=1/2$. 
In the thermodynamical limit, with $N\to\infty$ and  $L\to\infty$, while $\rho = N/L^2$ constant, we can estimate the critical point. 
%
%
%
Combining Eqs.~(\ref{eq:P_def}) and~(\ref{eq:P_c}) we obtain, for a given set 
of parameters $D$, $v_0$, $\sigma_0$, and $d$, the critical density $\rho_c$ 
above which a peak at large cluster sizes emerges:
\begin{eqnarray}\label{eq:critical_rho}
\rho_c = N^{1-\xi} \frac{D}{v_c \sigma_0 d^2} \,.  
\end{eqnarray}
From Eq.~(\ref{eq:critical_rho}) is clear that when $\xi<1$, $\rho_c$ diverges 
with the system size $N$. Only for $\xi=1$, $\rho_c$ is a finite quantity in 
the thermodynamical limit. 
Thus, for $\alpha = \beta$ there is a critical density $\rho_c$ above which the collective transition occur. 
On the contrary, for $\alpha < \beta$, $\rho_c \to \infty$ and any finite densitiy $\rho_0$ corresponds to the mono-disperse phase. 
Eq.~(\ref{eq:critical_rho}) predicts the critical noise intensity $\eta_c$ below which one expect the collective clustering 
 phase to occur  as $\eta_c \propto \sqrt{ \rho N^{\xi - 1}}$, from which follows that for $\xi = 1$, the critical noise scales as  $\eta_c \propto \sqrt{\rho}$. Interestingly, a similar scaling was reported for the onset of collective motion in simulations with the Vicsek model~\cite{vicsek1995, chate2008, peruani2011b}. 
%

%
\begin{figure*}
\centering
\resizebox{17cm}{!}{\rotatebox{0}{\includegraphics{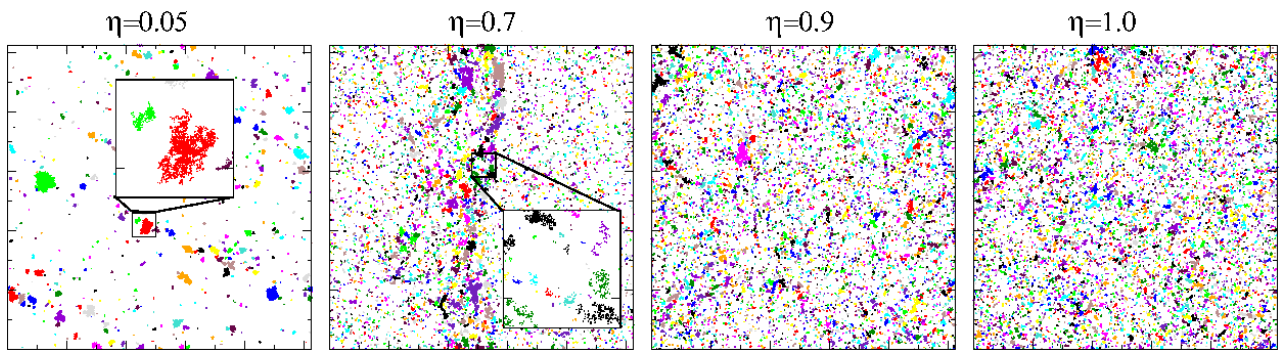}}}
\caption{Simulations snapshots of the VM for various noise intensity values. Particles belonging to the same cluster are labeled with the same color. As it can be seen in Fig. \ref{fig:vicsek}, for $\eta=1.0$ the system is disordered, while for $\eta<1$ there is preferred direction of motion. Though the snapshots for $\eta=1.0$ and $\eta=0.9$ look comparable, their CSD is remarkably different: an exponential distribution for the former, and a power-law for the later. Traveling bands in the VM are composed of many correlated clusters ($\eta=0.7$). Finally, for very low $\eta$  values most particles form part of very large clusters ($\eta=0.05$). } \label{fig:snapshots}
\end{figure*}

%
\begin{figure}
\centering
\resizebox{\columnwidth}{!}{\rotatebox{0}{\includegraphics{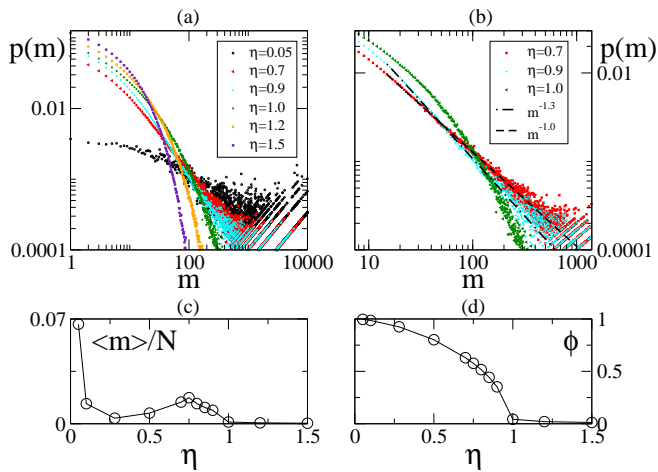}}}
\caption{Clustering in the VM. (a) shows the CSD for various $\eta$ values,  while (b) displays the scaling close to the critical point $\eta_c$. 
The average (normalized) cluster size $\langle m \rangle/N$ as function of $\eta$ is shown in (c), and the orientational order parameter $\phi$ vs. $\eta$ in (d). 
While the system is disordered, the CSD is dominated by an exponential tail. 
Below the critical point, the CSD is a power-law with an exponent $\gamma(\eta)$. 
Close to the critical point $\gamma = 1.3$ and as $\eta$ is decreased $\gamma$ approaches $1$. 
For very low $\eta$ values, the CSD decreases even slower than $1/m$  (see Fig. \ref{fig:snapshots}). } \label{fig:vicsek}
\end{figure}

\section{Clustering in the presence of global orientational order}\label{sec:globalOrder}
The simple set of Smoluchowski-type Eqs.~(\ref{rea_vicsek}) has been derived assuming that clusters move in an uncorrelated
fashion, as observed for instance, in bacterial experiments~\cite{peruani2012}.
However, several SPP models exhibit long-range orientational order and
macroscopic structures such as bands, which indicates that there are strong
correlations among cluster velocities. 
Thus, {\it a priori} we cannot expect Eq.~(\ref{rea_vicsek}) to describe 
quantitatively the cluster dynamics of such systems deep in phases with strong cluster-cluster correlations.   
Interestingly, it has been already observed that the Vicsek model (VM) in its (global) ordered phase displays  CSDs that can be power-law distributed, with $0.8 < \gamma < 1.3$, according to~\cite{chate2008, huepe2004}. 
Here, we carry out simulation in the Vicsek model (VM), i. e. Eqs.~(\ref{eq:mot_theta}) with a choice of $f=e^{i\theta^t_k}$.  Our simulation results 
confirm that indeed the VM exhibits power-law distributed CSDs with exponents in this range, and show that this occurs close to the well-known disorder-order transition of the VM. Furthermore, we find that the cluster statistics in the VM, below and close to the critical point -- i.e., when cluster-cluster correlations are weak or absent -- resembles that obtained with the kinetic clustering model. As expected, far away from the critical point and well in the ordered phase, cluster-cluster correlations induce effects that cannot be accounted by the kinetic clustering model. In short, the simulation data strongly suggests that close to the disorder-oder critical point,  the VM exhibits a transition to a  collective clustering phase as the one described with the kinetic model.  
  
%
 
The order-disorder transition in the VM is observed when the noise intensity $\eta$ is decreased below a critical value  $\eta^*$. 
Orientational order is characterized by $\phi = |\sum_j \exp(i \theta^{t}_{k})| /N$. 
In the disorder phase, the clusters cover homogeneously the space (Fig.~\ref{fig:snapshots}) and the CSD  is dominated by an exponential tail, see Fig.~\ref{fig:vicsek}. 
If $\eta$ is decreased below  $\eta^*$, velocity-velocity correlations among particles become important, $\phi$ increases, and clusters grow significantly in size.  
%
%
 Below the critical point $\rho^*$, the CSD is a power-law with an exponent in the range $[0.8, 1.3]$, which results also in a change of behavior of   $\langle m \rangle /N$ , as shown in Fig.~\ref{fig:vicsek}. 
 %
 %
 The power-law distributed CSD indicates that the system displays arbitrary large clusters. Indirectly, this also means that particle-particle correlations are long-ranged, i.e., arbitrarily large as the  cluster sizes. 
 Notice that this does not necessary imply the existence of cluster-cluster correlations. 
 Nevertheless, in the VM at low values of  the noise intensity  cluster-cluster correlations become evident. 
%
 %
 Arguably, due to these cluster-cluster correlations, we observe a clear deviation from what the theory predicts for $\eta << \eta^{*}$ . Moreover, in the VM, 
 $\gamma$ is a function of $\eta$. 
We observe that close to $\eta^*$, $\gamma \approx 1.3$, and as $\eta$ is decrease,  $\gamma$  approaches $1$.  
It is close to $\eta^*$ that we observe the emergence of a traveling band. Notice that this band is not a connected 
component, but a cloud of highly correlated moving clusters. We find that the CSD is power-law distributed in the band regime. 
For $\eta << \eta^{*}$ the system is highly ordered and clusters move roughly in the same direction,  which means that cluster collisions are less frequent. 
 On the other hand, low values of $\eta$ also imply that spreading of a cluster around its center of mass is also small. 
 In summary, at low noise values the cluster dynamics slows downs significantly in the VM, which results in a  non monotonous response of  $\langle m \rangle /N$ with $\eta$, Fig.~\ref{fig:vicsek}.  
 %
 %
Interestingly, we observe that at extremely low $\eta$ values the CSD decreases even slower than $1/m$ and $\langle m \rangle /N$ experiences a sharp increase.

\section{Concluding remarks}\label{sec:concludingRemarks}
We have argued that SPP systems exhibit for a finite system size    
two phases: a mono-disperse and a collective clustering  phase, characterized by a non monotonic CSD.  
Assuming that the moving directions of clusters are uncorrelated, Eq.~(\ref{rea_vicsek}) justifies the existence of these two phases. 
At the transition point, the CSD is a power-law characterized by an exponent $\gamma$  that depends, according the proposed kinetic clustering model,  
on the scaling with cluster size of the cluster cross-section -- characterized by an exponent $\alpha$ --  and cluster perimeter -- characterized by an exponent $\beta$. 
A systematic study of the $\alpha-\beta$ parameter space revealed that  $\gamma$ always falls in the range between $0.8$ and $1.5$, which is consistent with experimental observations~\cite{zhang2010, peruani2012} and simulations~\cite{peruani2006, yang2010}.  
However, we learn that in the thermodynamical limit, only for $\alpha=\beta$ the above mentioned transition occurs, while  
otherwise, the system remains in the mono-disperse phase. 
It is worth pointing out that for the special case $\alpha = \beta$ the kinetic model predicts a critical density value that is independent of the system size. Hence, the properties for the transition to nonequilibrium clustering  found in the kinetic model for intermediate values of $N$ hold true in the thermodynamic limit.  

%

In addition to the study of the kinetic clustering theory given by Eq.~(\ref{rea_vicsek}), we characterized the cluster statistics of the Vicsek model~\cite{vicsek1995}. 
Despite the fact that cluster-cluster correlations and global order  emerge in the simulations of the Vicsek model that are not treated in  Eq.~(\ref{rea_vicsek}),  
we found that a transition to a collective clustering phase is also present in the VM. Moreover, $\gamma$  falls in the expected range if simulations are performed with ferromagnetic alignment, as shown here and in~\cite{huepe2004, chate2008}, as well as with nematic alignment~\cite{peruani2010}. 
Notably, we found that  the CSD is power-law distributed for a range of noise intensity values $\eta$ close to the order-disorder critical point, with $\gamma$ function of $\eta$. 
In summary, the existence of  two clustering phases is found as well in SPP systems in the presence or absence of global orientational order. 

Our kinetic approach is complimentary to the ongoing effort of deriving coarse-grained nonlinear field equations for the description of active matter~\cite{toner1998, bertin2006, mishra2010, ihle2011, grossmann2012,  dunkel2012, gopinath2012, peshkov2012} (for a review, see~\cite{marchetti2012c, ramaswamy2010}) that are better suited to capture  large-scale structures or the emergence of global orientational order in large systems. 
In contrast, the observed clustering phenomena are typically studied in the absence of long-range order, on a small scale, and for intermediate system sizes. It is also worth noting, that the non-equilibrium clustering phase leads to 
 (apparent) giant number fluctuations~\cite{peruani2012} which often have similar properties as the ones predicted by Toner and Tu~\cite{toner1998} in the phase with long-range orientational order.

We expect a similar transition and cluster dynamics as the one described here  for all SPP systems where particle speed is not strongly affected by the local density. 
The presence of a density dependent speed can dramatically change the above given clustering picture, since now large size clusters are prone to slow down significantly. 
As result of this effect, the SPPs can form a single large cluster which coexists with a background gas of particles in the individual phase, as found in~\cite{peruani2011, marchetti2012a, marchetti2012b, mccandlish2012}.  
The cluster dynamics of these systems is likely to be related to an equilibrium-like phase separation, as suggested in~\cite{peruani2011}. 
%

Acknowledgement: We like to thank H. Chat{\'e}, F. Ginelli and J. Toner for useful discussions. MB acknowledges financial support by DFG through GRK 1558.

\bibliographystyle{apsrev}

\end{document}